\documentclass[twocolumn,aps]{revtex4}
\usepackage{graphicx}
\usepackage{color,graphicx}
\usepackage{dcolumn}
\usepackage{bm}
\usepackage{epsfig}
\usepackage{supertabular}
\usepackage[bookmarksopen]{hyperref}
\def\be {\begin{equation}}
\def\ee {\end{equation}}

\def\bea {\begin{eqnarray}}
\def\eea {\end{eqnarray}}

\begin{document}
\title{Effect of running coupling on photon emission from
quark gluon plasma}
\bigskip
\bigskip
\author{ Mahatsab Mandal}
\author{ Pradip Roy}
\affiliation{Saha Institute of Nuclear Physics, 1/AF Bidhannagar
Kolkata - 700064, India}
\author{Sukanya Mitra}
\author{Sourav Sarkar}
\affiliation{Variable Energy Cyclotron Centre, 1/AF Bidhannagar
Kolkata - 700064, India}

\begin{abstract}
We discuss the role of running coupling on the thermal photon yield
from quark gluon plasma. It is shown that the photon production rate 
from the partonic phase is considerably enhanced when
running coupling is considered with respect to a fixed value.
However, we show by explicit evaluation that
although this difference survives the space-time evolution the 
experimental data cannot distinguish between the two once the
hard contribution, which is an essential component of photon
production mechanism, is added. 
\end{abstract}

\maketitle

Detection of quark gluon plasma (QGP) in heavy ion collisions has received
significant attention in recent years.  
Among possible signals, electromagnetic probes are
one of the most promising tools to characterize the initial state of 
the collisions~\cite{jpr}. Owing to their weak coupling 
with the constituents of the system they tend to 
escape almost unscattered. In fact, photons (dileptons as well) 
can be used to determine the initial temperature, or equivalently the 
equilibration time.
By comparing the initial temperature with the transition 
temperature from lattice QCD, one can infer whether QGP is produced.

Photons are produced at various stages from (i) initial hard scattering of
partons, (ii)
scattering of charged particles in the thermal medium (QGP and hadronic matter) 
and (iii) from $\pi^0$ and $\eta^0$ decays.
If this decay contribution is 
subtracted from the total photon yield what is left is the direct 
(excess) photons. 
The thermal photon rate due to Compton and annihilation processes in a quark
gluon plasma has been calculated by several authors over the
last two decades~\cite{kap,brapi,aur1,aur2,arnold}.
In all these calculations the strong coupling, $\alpha_s$ is treated
as constant or function of temperature $T$. However, in case
of relativistic heavy ion collisions, apart from temperature 
there is also the momentum scale $k$. 
One has to take into account the case when $k \sim T$ and treat $\alpha_s$
to be function of both $k$ and $T$~\cite{prd75054031}. By
incorporating this fact it is shown that the energy loss is
a factor of $2-6$ more than the case when $\alpha_s$ is 
constant~\cite{prd75054031,gossiaux}. The energy loss calculation
using running coupling and reduced screening mass in \cite{gossiaux},
explains single electron $R_{AA}$ quite well.
It is the purpose of this brief report to treat the strong coupling
as running and apply it to the case of thermal photon production
from QGP. 

The lowest order processes for photon emission from QGP are the
Compton scattering ($q ({\bar q})\,g\,\rightarrow\,q ({\bar q})\,
\gamma$) and annihilation ($q\,{\bar q}\,\rightarrow\,g\,\gamma$)
process. The total cross-section diverges in the limit $t$
or $u \to 0$. These singularities have to be shielded by thermal
effects in order to obtain infra-red safe calculations. It has
been argued
in Ref.~\cite{kajruus} that the intermediate quark  acquires a thermal
mass in the medium, whereas the hard thermal loop (HTL) approach of
Ref.~\cite{brapi} shows that very soft modes are suppressed in a
medium providing a natural cut-off $k_c \sim gT$.
We assume that the singularities can be shielded by the
introduction of thermal masses for the participating partons.
The differential cross-sections for Compton and
annihilation processes are given by~\cite{wong},
\begin{eqnarray}
&&\frac{d\sigma(qg\to q\gamma)}{d\hat t}=\frac{1}{6}(\frac{e_q}{e})^2
\frac{8\pi
  \alpha_s \alpha_e}{(\hat s -m^2)^2}
(\frac{m^2}{\hat s -  m^2} +\frac{m^2}{\hat u -m^2})^2\nonumber\\
&&+(\frac{m^2}{\hat s -m^2}+\frac{m^2}{\hat u  -m^2})
- \frac{1}{4}(\frac{\hat s -m^2}{\hat u- m^2}+
\frac{\hat u - m^2}{\hat s-m^2})
\end{eqnarray}
and
\begin{eqnarray}
&&\frac{d\sigma(q \bar q\to g \gamma)}{d\hat t}=-\frac{4}{9}
(\frac{e_q}{e})^2
\frac{8\pi\alpha_s
  \alpha_e}{\hat s(\hat s-4m^2)}
(\frac{m^2}{\hat t-m^2}+\frac{m^2}{\hat u-m^2})^2
\nonumber\\
&&+(\frac{m^2}{\hat t-m^2}+\frac{m^2}{\hat u-m^2})
-\frac{1}{4}(\frac{\hat t-m^2}{\hat u-m^2}+\frac{\hat u-m^2}{\hat t-m^2})
\end{eqnarray}
where $m$ is the in-medium thermal quark mass,
$m^2\equiv {m_{th}}^2=2\pi \alpha_s T^2/3$, $\alpha_e$ and $\alpha_s$ are the
electromagnetic fine-structure and the strong
coupling constants respectively. The static photon rate in
$1+2\rightarrow 3+\gamma$ can be written as~\cite{jpr}
\begin{eqnarray}
\frac {dN}{d^4xd^2p_T dy}&=&\frac{\mathcal {N}_i}{(2\pi)^7 E}
\int d{\hat s} d{\hat t}
|\mathcal{M}_i|^2 \times\int dE_1 dE_2 \nonumber\\
&&\frac{f_1(E_1) f_2(E_2)(1+f_3(E_3))}{\sqrt{a{E_2}^2+2bE_2+c}}
\label{rate}
\end{eqnarray}
\rm where \nonumber
\begin{eqnarray}
a&=&-(\hat s+\hat t-{m_2}^2-{m_3}^2)^2 \nonumber\\
b&=&E_1(\hat s+\hat t-{m_2}^2-{m_3}^2)({m_2}^2-\hat t)\nonumber\\
&+&E[(\hat s+\hat t-{m_2}^2-{m_3}^2)\nonumber\\
&\times&(\hat s-{m_1}^2-{m_2}^2)-2{m_1}^2({m_2}^2-\hat t)]\nonumber\\
c&=&-{E_1}^2({m_2}^2-\hat t)^2\nonumber\\
&-&2E_1E[2{m_2}^2(\hat s+\hat t-{m_2}^2-{m_3}^2)
\nonumber\\
&-&({m_2}^2-\hat t)(\hat s-{m_1}^2-{m_2}^2)]\nonumber\\
&-&E^2[(\hat s-{m_1}^2-{m_2}^2)^2-4{m_1}^2{m_2}^2]\nonumber\\
&-&(\hat s+\hat t-{m_2}^2-{m_3}^2)({m_2}^2-\hat t)\nonumber\\
&\times&(\hat s-{m_1}^2-{m_2}^2)+{m_2}^2(\hat s+\hat t-{m_2}^2-{m_3}^2)^2
\nonumber\\ 
&+&{m_1}^2({m_1}^2-\hat t)^2\nonumber\\
E_{1,min}&=&\frac{\hat s+\hat t-{m_2}^2-{m_3}^2}{4E}
+\frac{E{m_1}^2}{\hat s+\hat t-{m_2}^2-{m_3}^2} \nonumber\\
E_{2,min}&=&\frac{E{m_2}^2}{{m_2}^2-\hat t}+\frac{{m_2}^2-\hat t}{4E},\ 
E_{2,max}=-\frac{b}{a}+\frac{\sqrt{b^2-ac}}{a}\nonumber~.
\end{eqnarray}
$\mathcal {M}_i$ represents
the amplitude for Compton or annihilation process. 
The overall degeneracy factor $\mathcal {N}_i=320/3$ and 20 for 
Compton and annihilation processes respectively involving $u$ and $d$ quarks.

As mentioned earlier, the infra-red cut-off is fixed by
plasma effects, where only the medium part is considered, completely
neglecting the vacuum contribution leading to ambiguity in the
calculation of cross-section at finite temperature QCD.
If the latter part is taken into account the strong
coupling should be running. Thus for any consistent calculation one has
to take this fact into consideration. We have in that case
$\alpha_s=\alpha_{s}(k,T)$  where $k=\sqrt{|t|}$.

Photons from thermal hadronic matter also make up an essential component of
the total photon yield from heavy ion collisions. These are emitted in
reactions between charged hadrons and in the radiative decays of unstable
hadrons~\cite{jpr,we1,we2}. In this work, we have used the amplitudes
of photon producing reactions involving the $\pi$, $\rho$, $\omega$, $\eta$, $K$
and $K^*$ mesons obtained in~\cite{turbide}.

The hard photon contribution can be calculated by perturbative QCD (pQCD).
In order to calculate photon production from reactions of the type $h_A\,h_B\,\rightarrow\,\gamma\,X$
(where $h_A, h_B$ refer to hadrons), we assume that the energy is such that
the  partonic degrees of freedom become relevant and they behave incoherently.
The cross-section for this process can then be written in terms of
elementary parton-parton cross-section multiplied by the partonic flux
which depends on the parton distribution functions~\cite{ctq6pdf}.
The energy scale for this to happen i.e. the factorization scale is denoted
by $Q^2$, the square of the momentum transfer of the reaction. 
Starting with two body scattering at the partonic level the differential 
cross-section for the reaction of above type can be written as~\cite{hadpo}
\begin{eqnarray}
\frac{d\sigma_{\gamma,\rm hard}}{d^2p_Tdy} &=& 
K\,\sum_{abc} \int_{x_a^{\rm min}}^{1}\,dx_a\,
G_{a/h_A}(x_a,Q^2)\, G_{b/h_B}(x_b,Q^2)\nonumber\\
&\times&\,\frac{2}{\pi}\,\frac{x_a x_b}{2x_a-x_T {e^y}}
\frac{d{\sigma}}{d{\hat t}}(ab\rightarrow \gamma c).
\label{eq8}
\end{eqnarray}
where, $x_T=2p_T/\sqrt{s}$ and the factor $K$ is introduced to
take into account the higher order effects. A few comments
about the $K$ factor is in order here. The cross-section in
the above expression is calculated perturbatively to leading order (LO)
in the strong coupling. 
In cases where the next-to-leading (NLO) order terms
are comparable to the LO terms the $K$ factor defined as NLO/LO is introduced 
in the LO computations to bring in the essence of the NLO terms. 
It has been shown in~\cite{kfac}
that $K$ depends on the choice of the momentum scale, the parton distribution functions
and the shadowing effect and its value lies between 2 - 3. In the present
calculation we take $K \sim 2.5$.

\begin{figure}
\begin{center}
\includegraphics[scale=0.32]{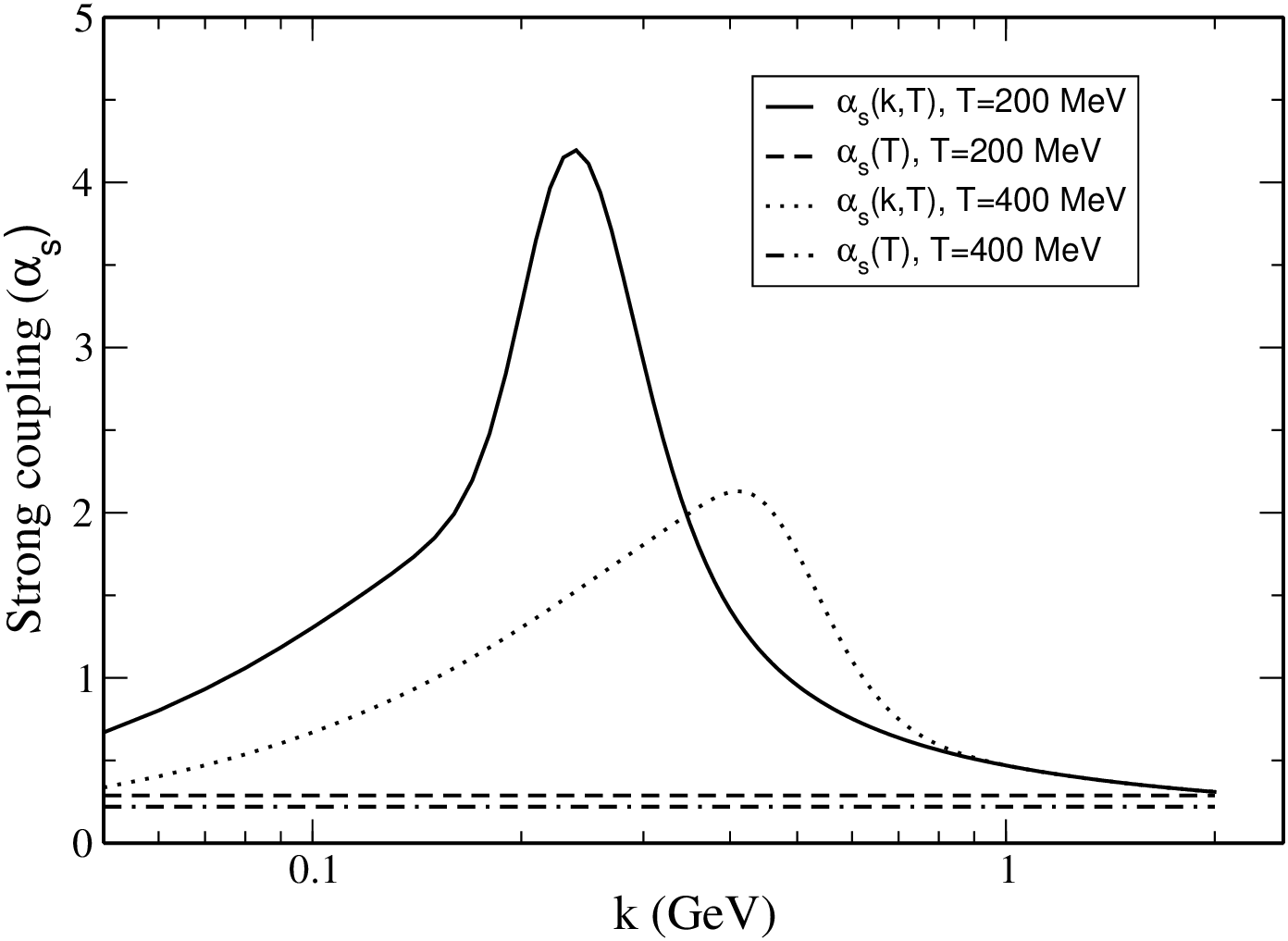}
\end{center}
\caption{Strong coupling as a function
of momentum scale at two different temperatures~\cite{prd75054031}.
}
\label{fig1}
\end{figure}

%
%
We also include photons from fragmentation process. This is accomplished by 
introducing the fragmentation function, $D_{\gamma /c}(z,Q^2)$, 
which when multiplied by $dz$ gives the probability for obtaining a photon from 
parton $c$, $z$ being the fractional momentum carried by the photon.
%
%
Once the photon production cross-section is obtained from hadron-hadron 
collision we can now determine the direct photon production rates due 
to hard scattering between partons from nucleus-nucleus collisions at 
relativistic energies. To do this we must note that the experimental data 
are given for a particular centrality. 
In order to take this into account we 
introduce the centrality parameter  
which depends on the maximum impact parameter $b_m$. The photon yield
from hard collisions is then calculated 
from the expression:
\begin{equation}
\frac{dN_{\rm AB}}{d^2p_T\,dy}(b_m) = {\cal R}(b_m)
\left[\frac{d\sigma_{\gamma,{\rm hard}}}{d^2p_T\,dy}+  
\frac{d\sigma_{\gamma,{\rm frag}}}{d^2p_T\,dy}\right]  
\label{eq18}
\end{equation}
where
${\cal R}(b_m) \equiv \langle ABT_{\rm AB} \rangle=
\frac{\int_0^{b_m}\,d^2{\bf b}\,AB\,T_{\rm AB}(b)}{\int_0^{b_m} d{\bf b}\,
\left(\frac{}{}1-[1-T_{\rm AB}(b)\sigma_{NN}^{in}]^{AB}
\right)} $
%
%
%
%
and
\(T_{\rm AB}({\bf b}) = \int\,d^2{\bf s}\,T_{\rm A}({\bf s})\,T_{\rm B}(\bf
b-s)\)
is the nuclear overlap function. For 0 - 10 \% centrality we obtain
${\cal R} \sim 21.7 mb^{-1}$.
Before going to the numerical evaluation of the static photon rate
we plot the running coupling in Fig.~(\ref{fig1}) where the
parametrization for $\alpha_s(k,T)$ is taken from Ref.~\cite{prd75054031}. 
It is seen that the value
of the coupling is largest when $k \sim T$. For $k >> T$ it agrees
well with the temperature dependent coupling. These features of
$\alpha_s$ have important consequence on the photon production rate
as we shall see below.

\begin{figure}
\begin{center}
\includegraphics[scale=0.31]{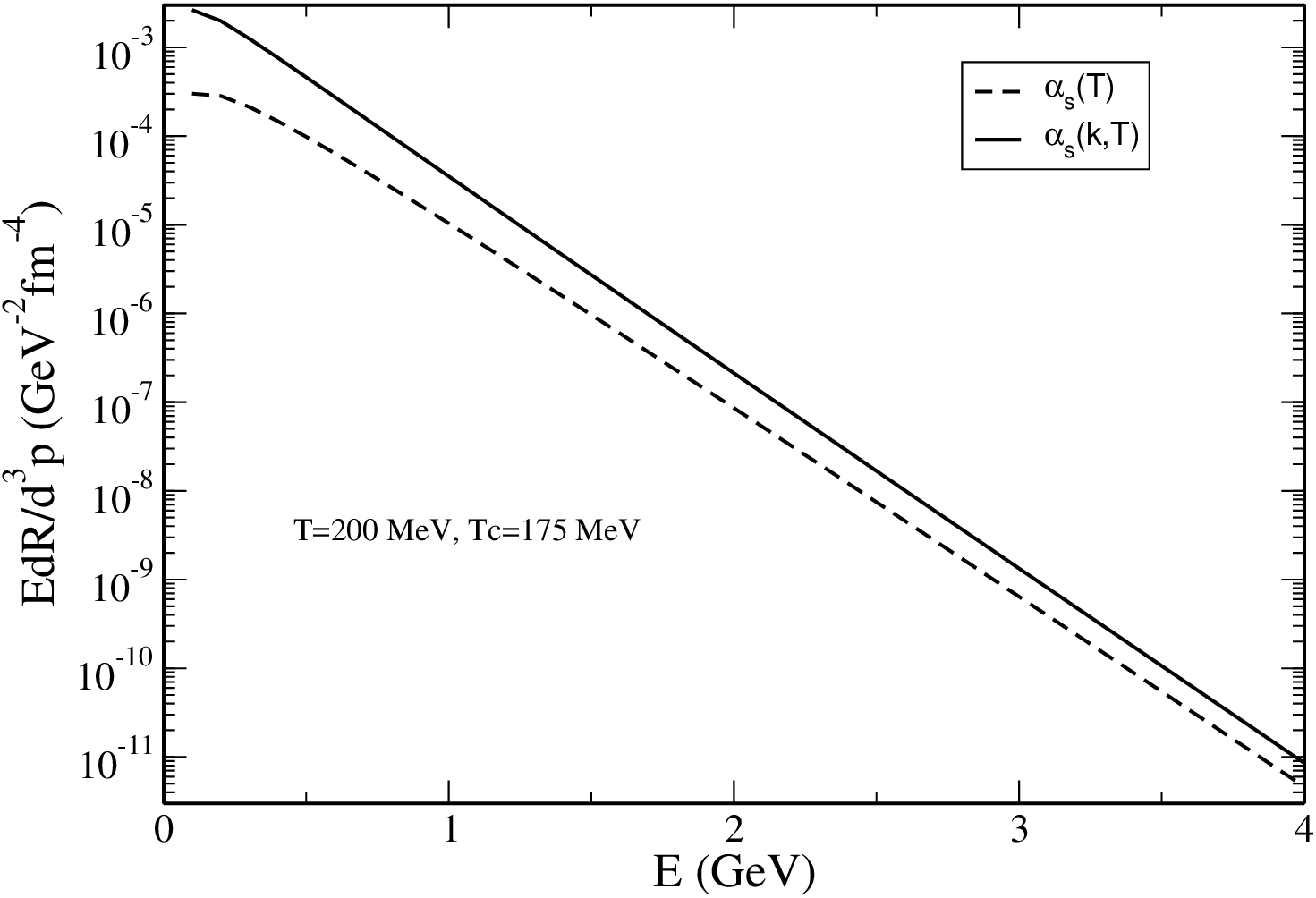}
\end{center}
\caption{Static photon rate with and without the running coupling.
}
\label{fig2}
\end{figure}

The static photon rate is obtained from Eq.(3) using
the running coupling. For $T=200$ MeV 
the rates are shown in Fig.~(\ref{fig2}).
The photon emission rate is enhanced by a factor of 1.7 - 6 compared
to the case when the momentum dependence of the strong coupling is neglected.

Photons are produced at all stages of the collision and so it is necessary to
integrate the emission rates over the space-time volume from
creation to freeze-out. We assume  
that quark gluon plasma having a temperature $T_i$
is produced at an initial time $\tau_i$. Hydrodynamic expansion and cooling follows up to
a temperature $T_c$ where QGP crosses over to a hadronic gas. Subsequent
cooling leads to freeze-out of the fluid element into observable hadrons.
In the present work the fireball is taken to undergo an azimuthally symmetric
transverse expansion along with a boost invariant longitudinal expansion.
The local temperature of the fluid element and the associated flow
velocity as a function of the radial coordinate and proper time is obtained
by solving the the energy momentum conservation equation
$\partial_\mu\,T^{\mu \nu}=0$
where $T^{\mu \nu}=(\epsilon+P)u^{\mu}u^{\nu}\,+\,g^{\mu \nu}P$ is the energy
momentum tensor for ideal fluid.  This set
of equations are closed with the Equation Of State (EoS); 
typically a functional relation between the pressure $P$ and 
the energy density $\epsilon$. It is a crucial input which essentially controls
the profile of expansion of the fireball. To minimize model dependencies we take
the EoS from the lattice calculations of the Wuppertal-Budapest
Collaboration~\cite{borsanyi}. 

The initial temperature is related 
to the experimentally measured hadron multiplicity
through entropy conservation~\cite{hwa} as
\(T_i^3(b_m)\tau_i=\frac{2\pi^4}{45\zeta(3)\pi\,R_A^2 4a_k}
\langle\frac{dN}{dy}(b_m)\rangle
\)
where $\langle dN/dy(b_m)\rangle$ is the hadron (predominantly pions) 
multiplicity
for a given centrality class with maximum impact parameter $b_m$, 
$R_A$ is the transverse dimension of the system and
$a_k$ is the degeneracy of the system created.
The hadron multiplicity  resulting from $Au + Au$ collisions
is related to that from $pp$ collision at a given
impact parameter and collision energy through the relation 
$\langle \frac{dN}{dy}(b_m)\rangle=\left[(1-x)\langle N_{part}(b_m)\rangle/2
+x\langle N_{coll}(b_m)\rangle\right]\frac{dN_{pp}}{dy}$
where $x$ is the fraction of hard collisions.
$\langle N_{part}\rangle$
is the average number of participants and  $\langle N_{coll}\rangle$ is the 
average number of collisions
evaluated by using Glauber model.
$dN_{pp}^{ch}/dy= 2.5-0.25\ln s+0.023\ln^2s$ 
is the multiplicity of the produced hadrons
in $pp$ collisions at centre of mass energy, $\sqrt{s}$~\cite{KN}.
We have assumed that $20\%$ hard (i.e. $x=0.20$ ) and
$80\%$ soft collisions are responsible for initial entropy production. 
Considering 0 - 10 \% centrality we get
$T_i=400$ MeV for $\tau_i=0.2$ fm/c.
One also requires the initial energy density and  radial velocity profiles 
which are taken respectively as~\cite{hvg}
$\epsilon(\tau_i,r)=\frac{\epsilon_0}{1+e^{(r-R_A)/\delta}}$ and 
$v(\tau_i,r) = v_0\left[1-\frac{1}{1+e^{(r-R_A)/\delta}}\right]$, 
where $\delta$ ($\sim 0.5$ fm) is a parameter, known as the surface 
thickness. As discussed in~\cite{hvg}, this choice of the initial fluid velocity profile is
motivated by the fact that for a physical system the initial fluid velocity is
zero inside the matter which approaches a value $v_0$ which is of the order
of a typical particle transverse velocity in the diffuse region
outside the matter distribution.

\begin{figure}
\begin{center}
\includegraphics[scale=0.33]{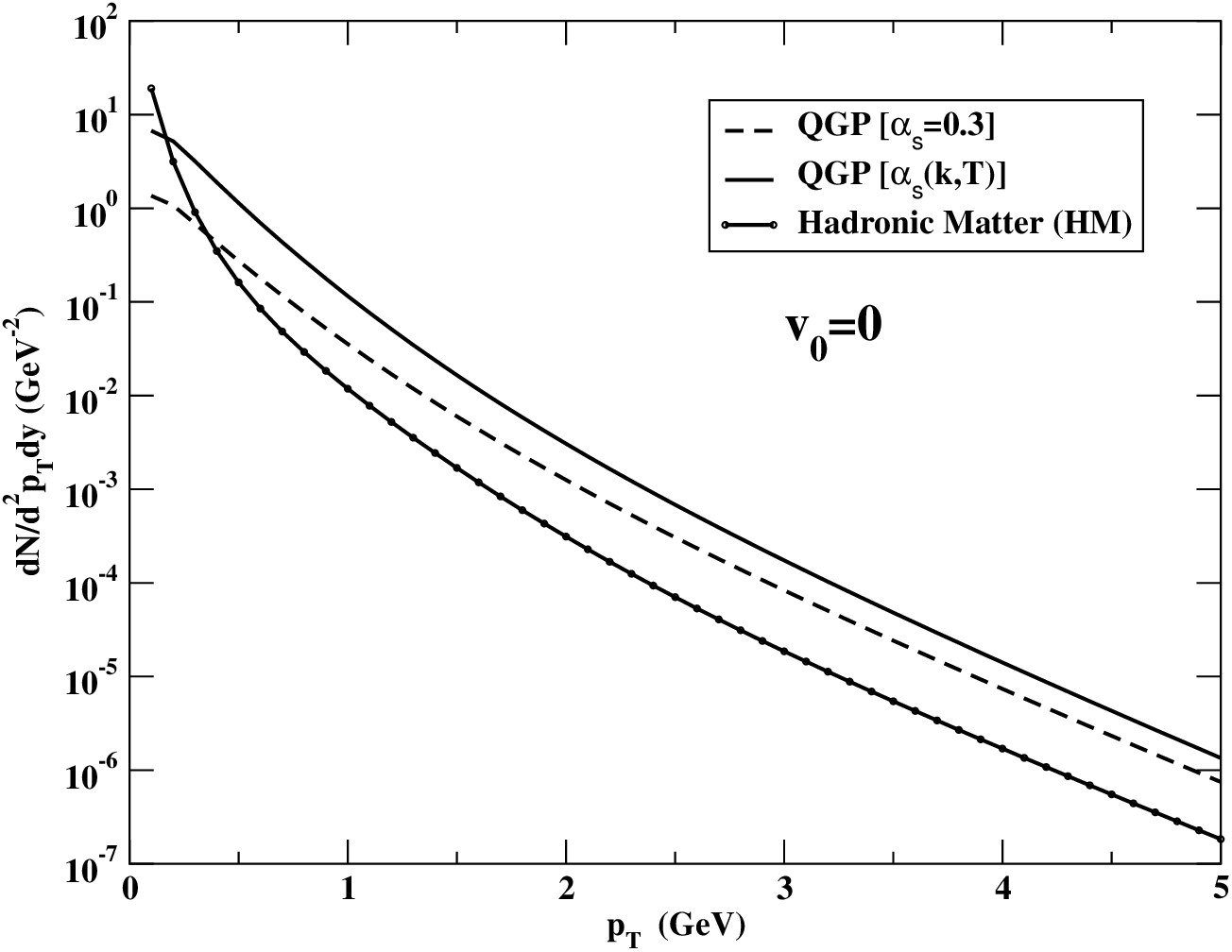}
\includegraphics[scale=0.30]{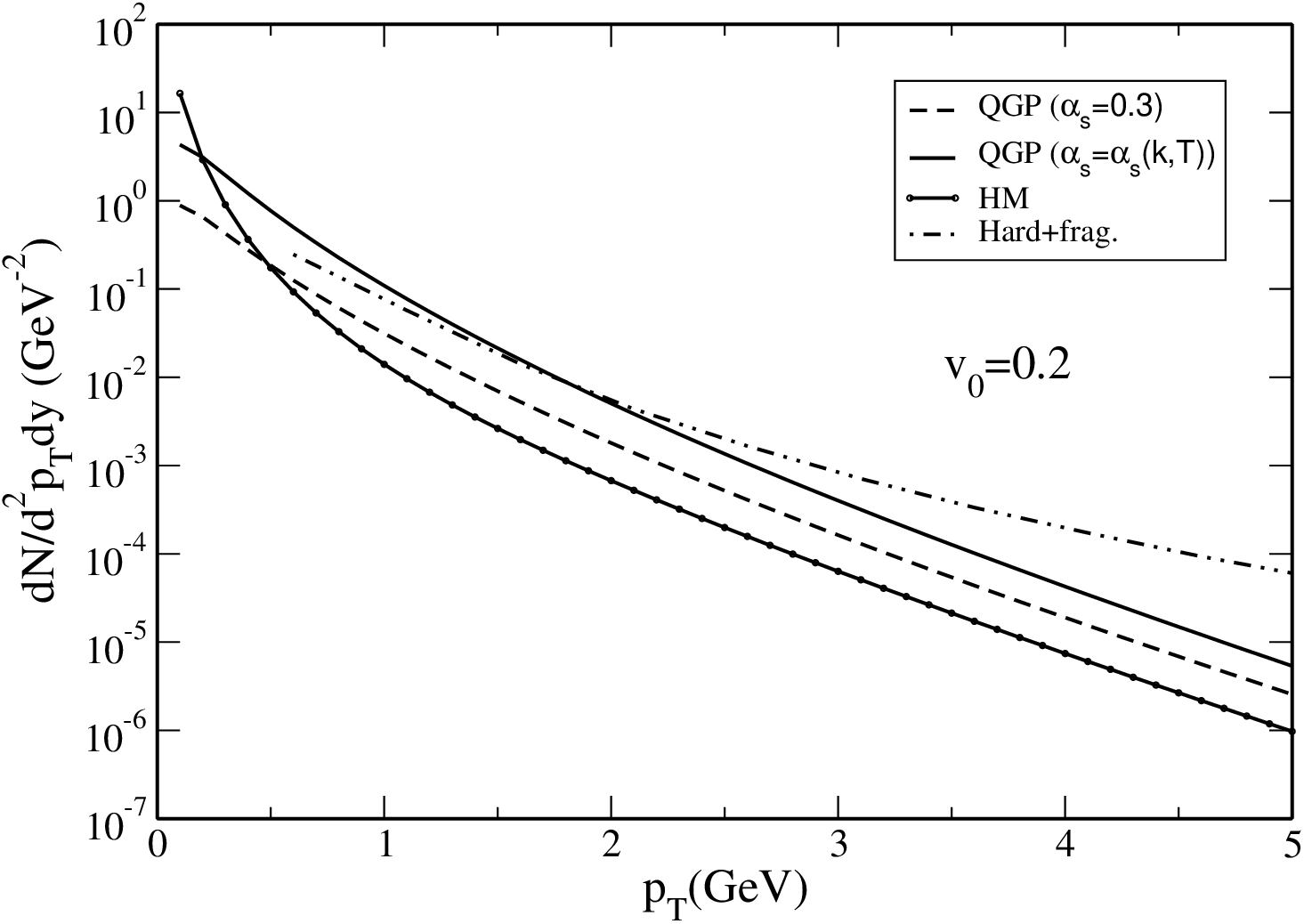}
\end{center}
\caption{Thermal photon $p_T$ distributions for $T_i$ = 400 MeV and 
$\tau_i$ = 0.2 fm/c at central rapidity. 
Upper panel corresponds to  $v_0$ = 0. 
Individual contribution from hard and fragmentation photons
has been shown in the lower panel for $v_0$ = 0.2. 
}
\label{fig2a}
\end{figure}

The other inputs are the transition temperature $T_c$ which 
is taken as  175 MeV as 
obtained from lattice QCD~\cite{katz,cheng} and 
the freeze-out temperature, $T_f$ which is taken to be 120 MeV.
We now plot the thermal photon yield from both QGP and hot hadronic matter in 
Fig.(\ref{fig2a}) for $v_0=0$ and 0.2 in the upper and lower panels.
At very low $p_T \sim 0.5$ GeV, the contribution from
hadronic matter dominates. Beyond that the QGP contribution starts to take
over. Interestingly, the effect of running coupling on the
thermal photon production from QGP does survive the space
time evolution and continues to be discernible in this $p_T$ domain.
The effect of non-zero initial velocity ($v_0$) is also
visible in the upward shift of the spectra at higher $p_T$ in the lower panel 
compared to the upper. Moreover, the relative separation between the
QGP contributions with and without the running coupling appears to be 
independent of
the space-time evolution scenarios corresponding to $v_0=0$ and 0.2.
To assess their relative importance in comparison to the
thermal yield we also show in the lower panel the contribution from hard QCD
photons and photons produced from jet fragmentation. As
shown by the dash-dotted line this contribution clearly 
dominates the photon yield beyond about 2 GeV. Below that the hard and
fragmentation contributions are almost similar to QGP contribution.


\begin{figure}
\begin{center}
\includegraphics[scale=0.34]{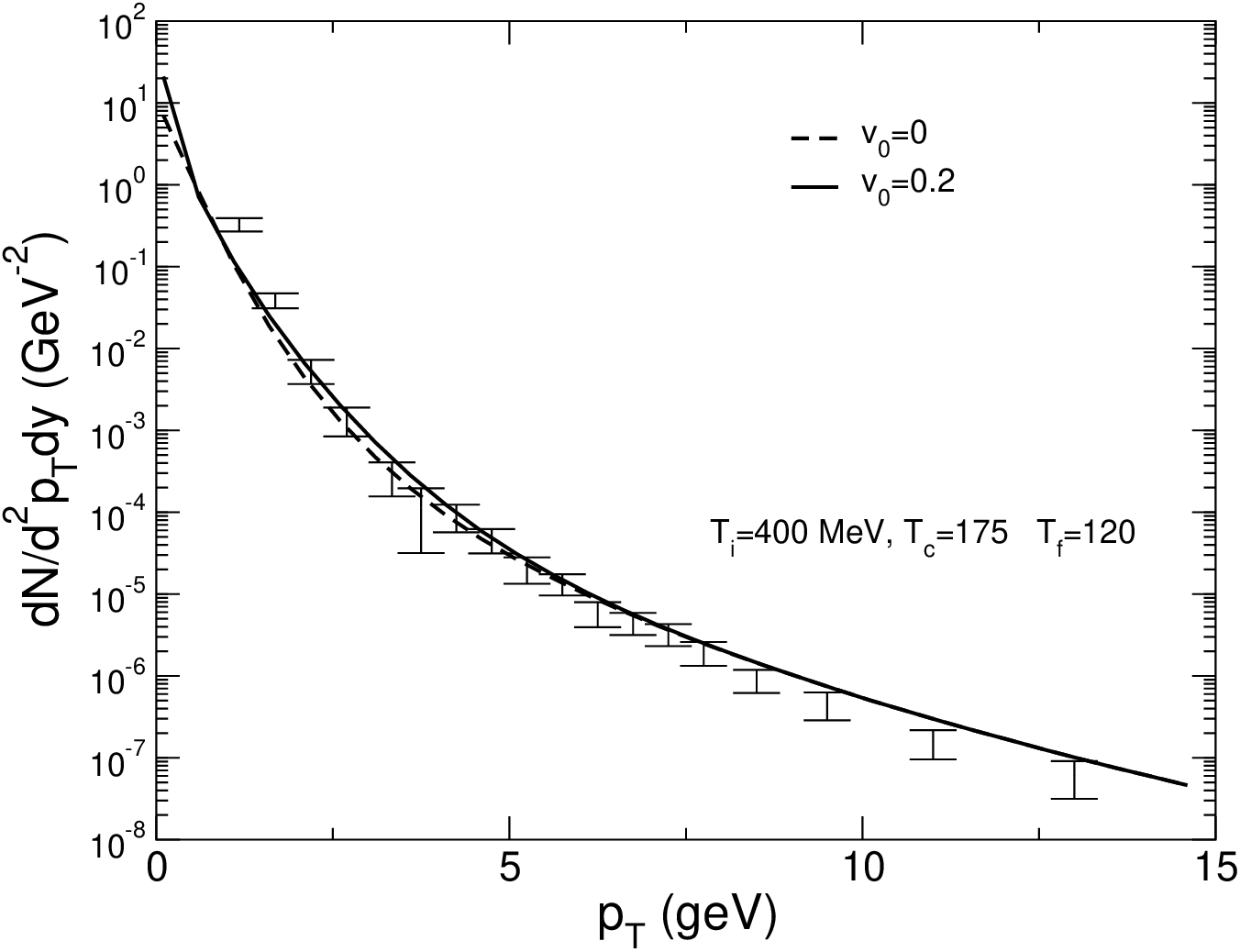}
\end{center}
\caption{Photon $p_t$ distributions for $T_i=400$ MeV
and $\tau_i=0.2$ fm/c at central rapidity.
The data (for $|y|\leq 0.35$)
are taken from Ref.~\protect\cite{phenix}.} 
\label{fig3}
\end{figure}

We now compare the total yield with the direct photon data from Au+Au 
collisions at RHIC
obtained by PHENIX~\cite{phenix} in fig.~(\ref{fig3}).
It is observed that the data is best reproduced by assuming a small initial
velocity of the order of 0.2 (solid line) compared to $v_0=0$ (dashed line).
 However, the curves for $\alpha_s=\alpha_s(T)$ and $\alpha_s=\alpha_s(k,T)$
have merged with each other implying that the observed difference seen
in the thermal photon contribution
for the two cases has been washed away once the hard and
fragmentation contributions are added. Such a result
can be understood once we realize that the contribution at a given value of the
transverse momentum, especially up to 2-3 GeV/c is a superposition of
contributions from QGP at temperatures from $T_i$ to $T_c$, hadronic matter from
$T_c$ to $T_f$ as well as from hard scatterings. 
Although the QGP contribution clearly dominates for $p_T > 0.5$ GeV  
observation of  momentum-dependence of the strong coupling in the 
transverse momentum spectra of single photons does not appear to be feasible 
as it is overshadowed by the contributions coming from initial hard 
collisions.  


To summarize, we have calculated the static photon rate from QGP due to Compton 
and annihilation processes using the temperature and momentum dependent strong 
coupling. The rate is then contrasted with the case where $\alpha_s$
depends only on the temperature of the system. It is found that the
static photon rate enhances significantly if the
running coupling is used. We then perform a space-time evolution
using relativistic hydrodynamics with initial
conditions for Au+Au collisions at 200 GeV/n at RHIC.
 The significant difference in the yields does survive the space time evolution
and can be observed in the {\it thermal} photon spectra.
However, due to the large
contributions coming from the initial hard collisions and jet fragmentation
the single photon data from PHENIX cannot distinguish between
the scenarios with and without momentum dependence of the running coupling.

\end{document}